\documentclass[preprint,aps,tightenlines,showpacs,amsmath,amssymb]{revtex4-1}
\usepackage[dvips]{graphicx}
\usepackage{subfigure}
\usepackage{color}
\usepackage{epsfig}
\usepackage{indentfirst}
\usepackage{multirow}
\usepackage{amsmath}
\usepackage{amssymb}
\usepackage{bm}
\usepackage{fancyhdr}
\usepackage{enumerate}
\usepackage[bookmarksnumbered,pdfstartview=FitH,colorlinks=true,menucolor=black,linkcolor=blue]{hyperref}

\newcommand{\psp}{\psi(3686)}

\newcommand{\jpsi}{J/\psi}

\newcommand{\beq}{\begin{equation}}
\newcommand{\eeq}{\end{equation}}
\newcommand{\bitm}{\begin{itemize}}
\newcommand{\eitm}{\end{itemize}}

\newcommand{\gevcc}{\mathrm{GeV}/\mathrm{c}^2}
\newcommand{\gevc}{\mathrm{GeV}/\mathrm{c}}
\newcommand{\gev}{\mathrm{GeV}}
\newcommand{\pbv}{\mathrm{pb}^{-1}}
\usepackage{lineno}

\begin{document}
\pagenumbering{Roman}

\title{%
\boldmath  Measurement of the branching fraction for $\psp\to\omega K^+ K^-$
}

\date{\today version:3}

\author{\small
M.~Ablikim$^{1}$, M.~N.~Achasov$^{8,a}$, X.~C.~Ai$^{1}$, O.~Albayrak$^{4}$, M.~Albrecht$^{3}$, D.~J.~Ambrose$^{41}$, F.~F.~An$^{1}$, Q.~An$^{42}$, J.~Z.~Bai$^{1}$, R.~Baldini Ferroli$^{19A}$
% ...
\\
\vspace{0.2cm}
(BESIII Collaboration)\\
\vspace{0.2cm} {\it
$^{1}$ Institute of High Energy Physics, Beijing 100049, People's Republic of China\\
% ...
$^{3}$ Bochum Ruhr-University, D-44780 Bochum, Germany\\
$^{4}$ Carnegie Mellon University, Pittsburgh, Pennsylvania 15213, USA\\
% ...
$^{8}$ G.I. Budker Institute of Nuclear Physics SB RAS (BINP), Novosibirsk 630090, Russia\\
% ...
$^{19}$ (A)INFN Laboratori Nazionali di Frascati, I-00044, Frascati, Italy; (B)INFN and University of Perugia, I-06100, Perugia, Italy\\
% ...
$^{41}$ University of Rochester, Rochester, New York 14627, USA\\
$^{42}$ University of Science and Technology of China, Hefei 230026, People's Republic of China\\
% ...
$^{a}$ Also at the Novosibirsk State University, Novosibirsk, 630090, Russia\\
% ...
}}

\vspace{20mm}

\begin{abstract}
With $1.06\times 10^8$ $\psi(3686)$ events collected with the BESIII
detector, the branching fraction of $\psi(3686) \to \omega K^+ K^-$ is
measured to be $(1.54 \pm 0.04 \pm 0.11) \times 10^{-4}$.  This is the
most precise result to date, due to the largest $\psi(3686)$ sample,
improved signal reconstruction efficiency, good simulation of the
detector performance, and a more accurate knowledge of the continuum
contribution.  Using the branching fraction of $J/\psi \to \omega K^+
K^-$, the ratio $\mathcal{B}(\psi(3868) \to K^+K^-) /
\mathcal{B}(J/\psi \to K^+K^-)$ is determined to be $(18.4 \pm
3.7)\,\%$. This constitutes a significantly improved test of the
$12\,\%$ rule, with the uncertainty now dominated by the $J/\psi$
branching fraction.
%
\end{abstract}


\maketitle
\pagenumbering{arabic}
\graphicspath{{Figures/}}

\section{Introduction}
\label{sec:introduction}

Since the experimental discovery of the
charmonium state $\jpsi$~\cite{Aubert:1974js} in 1974, four
decades have passed and much experimental and theoretical progress has
been achieved. However, puzzles still exist, and the ``$\rho \pi$ puzzle''
is one of the most famous.
From perturbative QCD (pQCD), it is expected that both $J/\psi$ and
$\psp$ decaying into light hadrons are dominated by the annihilation
of $c\bar c$ into three gluons, with widths proportional to the square
of the wave functions at the origin $|\Psi(0)|^2$~\cite{TA}. This
yields the pQCD ``12$\%$ rule'':
$$
Q_h=\frac{\mathcal{B}_{\psp\to h}}{\mathcal{B}_{J/\psi\to h}}
\approx \frac{\mathcal{B}_{\psp\to e^+e^-}}{\mathcal{B}_{J/\psi\to
e^+e^-}} =12.7\% \ .
$$
However, violations of this rule have been found in
experiments, and the first and most famous one was observed in
the $\rho\pi$ decay mode by Mark II~\cite{Franklin:1983ve},
which is now known under the name ``$\rho\pi$ puzzle"


Various decay channels have been studied to test the $12\%$ rule, and
for different decay modes the experimental ratios can be larger than,
smaller than, or consistent with $12\%$. Many possible mechanisms for
the violation of the $12\%$ rule have been proposed, but none of them
provide a universally satisfactory explanation at present. A
review can be found in Ref.~\cite{Asner:2008nq}. More experimental
studies of the branching fractions of different $J/\psi$ and
$\psi(3686)$ decay modes are helpful to understand this puzzle. At
present, most measurements consider two-body decays; studies of
three- or more-body decays of $J/\psi$ and $\psp$ will provide
complementary information of the decay mechanism and may shed light on
the $\rho \pi$ puzzle.

The world average value of the branching fraction for $\psp\to\omega K^+K^-$ is
$(1.85\pm 0.25)\times 10^{-4}$ with an error greater than
$13\%$~\cite{Beringer:1900zz}.  The world averaged branching fraction of $J/\psi\to\omega K\bar K$
~\cite{Beringer:1900zz} is $(1.70\pm0.32)\times10^{-3}$, so
$\mathcal{B}(J/\psi\to\omega K^+K^-)$ is determined to be $(0.85\pm 0.16)\times 10^{-3}$
from isospin symmetry. Then the ratio $Q \equiv \mathcal{B}_{\psp \to \omega K^+ K^-}/
\mathcal{B}_{J/\psi \to \omega K^+ K^-} = (21.8\pm 5.0)\%$. While its mean value disagrees
with the $12\%$ rule, it is still marginally consistent with $12.7\%$ considering the
large uncertainty.

In this paper, we measure the branching fraction of $\psi(3686) \to
\omega K^+K^-$ using $1.06 \times 10^8$ $\psi(3686)$ events collected
with the BESIII detector at BEPCII; furthermore, $44.49~\pbv$\cite{BESIII:2013iaa} of
$e^+e^-$ data collected at $3.65~\gev$ is used to determine
the continuum contribution. This new and more precise result will be
used to determine the ratio $Q$ for this decay channel.

\section{Detector and Monte Carlo simulation}
The Beijing Electron Positron Collider (BEPCII)~\cite{Ablikim:2009aa} is a double-ring
$e^+e^-$ collider designed to provide a peak luminosity of $10^{33}\
\mathrm{cm}^{-2}{\mathrm s}^{-1}$, and the BESIII~\cite{Ablikim:2009aa} detector is a
general-purpose detector designed to take advantage of this high luminosity.  Momenta of
charged particles are measured in a 1 T magnetic field with a resolution $0.5\%$ at
$1~\gevc$ in the helium-based main drift chamber (MDC), and the energy loss ($dE/dx$) is
also measured with a resolution better than $6\%$.  The energies and positions of neutral
tracks are measured in the electromagnetic calorimeter (EMC) composed of $6240$ CsI (Tl)
crystals. The energy resolution of $1.0~\gev$ photons is $2.5\%$ in the barrel and $5.0\%$
in the end-cap regions; the position resolution is $6$ mm in the barrel and $9$ mm in the
end-cap regions. In addition to $dE/dx$, a time-of-flight system (TOF) contributes to
particle identification with a time resolution of $80$ ps in the barrel and $110$ ps in
the end-cap regions. The muon system, interspersed in the steel plates of the magnetic
flux return yoke of the solenoid magnet, consists of $1272\ {\mathrm m}^2$ of resistive
plate chambers (RPCs) in $9$ barrel and $8$ end-cap layers, which provide a position resolution
of $2\ \mathrm{cm}$.



The optimization of the event selection and the estimation of physics
backgrounds are performed using Monte Carlo (MC) simulated data
samples. The {\sc geant4}-based~\cite{Agostinelli:2002hh} simulation software {\sc
boost}~\cite{boost} includes the geometric and material description of
the BESIII detectors and the detector response and digitization
models, as well as the tracking of the detector running conditions and
performance. In the MC of inclusive $\psp$ decays, the production of the $\psp$ resonance is simulated by
the MC event generator {\sc kkmc}~\cite{Jadach:1999vf}; the known decay modes
are generated by {\sc besevtgen}~\cite{Ping:2008zz} with branching fractions
set at world average values~\cite{Beringer:1900zz}, while the remaining unknown
decay modes are modeled by {\sc lundcharm}~\cite{Chen:2000tv}.
In the exclusive MC, the $\psp\to\omega K^+K^-$ and $\omega\to\pi^+\pi^-\pi^0$ decays are generated with a new data-driven generator
based on EvtGen~\cite{evtgen}.

\section{Event selection and data analysis}
Each $\psp\to\omega K^+K^-$, $\omega \to \pi^+ \pi^- \pi^0$ candidate
has four good charged tracks with zero net charge and at least
two good photon candidates. A good charged track is required to
satisfy track fitting, and  pass within 10 cm
of the interaction point in the beam direction and within
1 cm in the plane perpendicular to the beam.
Furthermore, it is required to lie within
the angular coverage of the MDC, i.e. $|\cos\theta|<0.93$ in the
laboratory frame, where $\theta$ is the polar angle.

For photon candidates, the shower energy should be greater than $25$
MeV in the barrel region and $50$ MeV in the end-cap regions, where
the barrel is defined as $|\cos\theta| < 0.8$ and the end-cap regions
as $0.86 < |\cos\theta| < 0.92$.
Also the average time of the hit EMC crystals with respect to the event start time
should be
between $0$ and $700$ ns to suppress electronic noise and background
hits. The angle between the direction of a photon candidate and any
charged track is required to be greater than $20^\circ$ to avoid
showers caused by charged tracks.

The TOF and $dE/dx$ information are combined for each charged track to
calculate the particle identification probability ($P_i$ with $i=\pi,\ K$) of each particle
type hypothesis. For a pion candidate, $P_\pi>0.001$ and $P_\pi>P_K$
are required, while for a kaon candidate $P_K>0.001$ and $P_K>P_\pi$
are required.

A vertex fit is performed assuming all charged tracks are from the IP.
A four-constraint (4C) energy-
momentum conserving kinematic fit is performed.
If there are more than two
photon candidates, we loop over all possible combinations, and the
combination with the minimum 4C $\chi^2$ is kept for further analysis.  The
invariant mass of the photon pair is required to be in the range
$0.11< M_{\gamma\gamma} < 0.15$ GeV/$c^2$. Then a 5C
kinematic fit is performed with the invariant mass of the two photons
constrained to the mass of $\pi^0$, and $\chi^2 < 90$ is
required, which is based on the optimization of the figure of merit
(FOM), $\mathrm{FOM} \equiv N_{sig}/\sqrt{N_{sig}+N_{bg}}$, where
$N_{sig}$ and $N_{bg}$ are the numbers of signal and background events
estimated by the inclusive MC, respectively.

\begin{figure}
\includegraphics[width=5.0in]{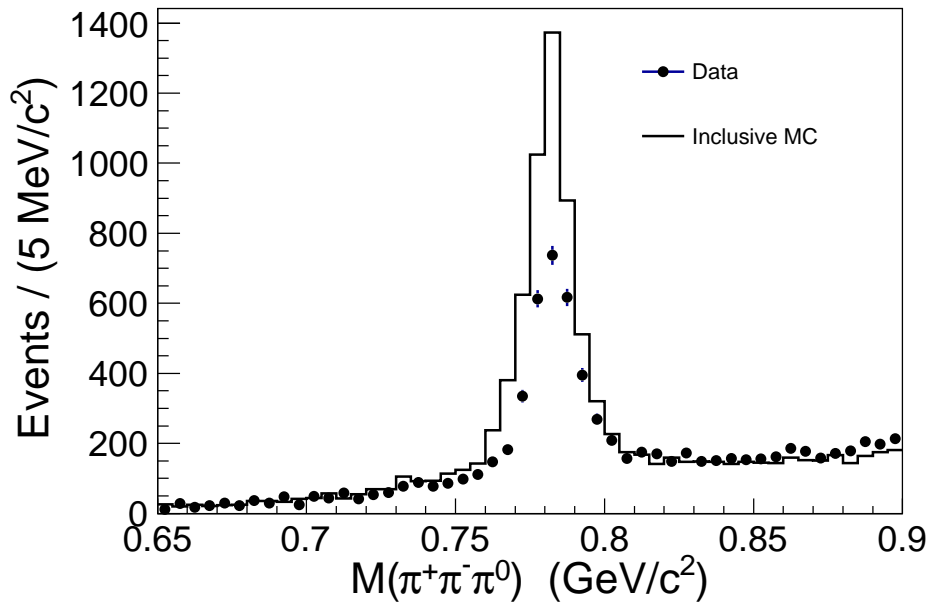}
\caption{Comparison of the $\pi^+ \pi^- \pi^0$ invariant mass
  distributions between the inclusive MC simulation and data at $3.686~\gev$.
  The histogram is inclusive MC that is normalized to the luminosity of data, while points with error bars are
  data.}
\label{fig:07}
\end{figure}

After all above mentioned selection criteria are applied, the $\pi^+\pi^-\pi^0$
invariant mass distributions of inclusive MC events and data are shown
in Fig.~\ref{fig:07}, in which the points with error bars are data
and the histogram is inclusive MC.
The inclusive MC sample, which contains the same number of events as the $\psp$ data
and uses the world average $\psp \to \omega K^+ K^-$ branching fraction~\cite{Beringer:1900zz}, \
has more signal events than the data in the region of $0.772< m_{\pi^+\pi^-\pi^0} < 0.792~\gevcc$.
The selected events with final states $\omega K^+K^-$ in the inclusive MC are mainly from three decay channels: $\psp\to\omega
K^+K^- \ (\mathrm{direct})$, $\psp\to K_1(1270)K,\ K_1(1270)\to\omega
K$ and $\psp\to \omega f_2(1270),\ f_2(1270)\to K^+K^-$.
And for the distribution of the invariant mass of $\pi^+\pi^-\pi^0$, no peaking background of the $\omega$ signal
is found in the fit region ($0.65< m_{\pi^+\pi^-\pi^0} < 0.9~\gevcc$).
The background simulation from the
inclusive MC is reliable as the side-band regions, defined as
$0.732< m_{\pi^+\pi^-\pi^0} < 0.752~\gevcc$ and
$0.812< m_{\pi^+\pi^-\pi^0} < 0.832~\gevcc$, match
well with data.

Due to the soft transition photon, there is also a possible peaking
background from $\psi(3686)\to \gamma \eta_c(2S)$, $\eta_c(2S)\to
\omega K^+ K^-$, which is not simulated in the inclusive MC sample.
It is studied based on an exclusive MC sample of $\psi(3686)\to \gamma
\eta_c(2S)$, $\eta_c(2S)\to \omega K^+ K^-$ assuming the branching
fraction of $\eta_c(2S)\to \omega K^+ K^-$ is $10^{-3}$ and taking
other branching fractions from the world average value~\cite{Beringer:1900zz}.
The contribution from this process is very small, about $0.1\%$ of the
observed $\omega K^+K^-$ candidates, and it is ignored.

To determine the signal efficiency, BODY3, a new data-driven generator based on
EvtGen~\cite{evtgen}, is used. BODY3 was developed to simulate contributions from
different intermediate states
or direct production in data for a given three-body final state.
First a MC sample is generated with phase space (PHSP) to determine efficiencies over the whole allowed kinematic region.
Next a Dalitz plot and two angular distributions, corrected for efficiency, are used to determine the probability of an event configuration generated randomly by Monte Carlo.
For our case, the Dalitz plot of the square of the $\omega K^+$ mass versus the square of the $\omega K^-$ mass and the angular distributions of the $K^+$ and $K^-$ in the $\psp$ CMS from data are used.

\begin{figure}
\includegraphics[width=6.0in]{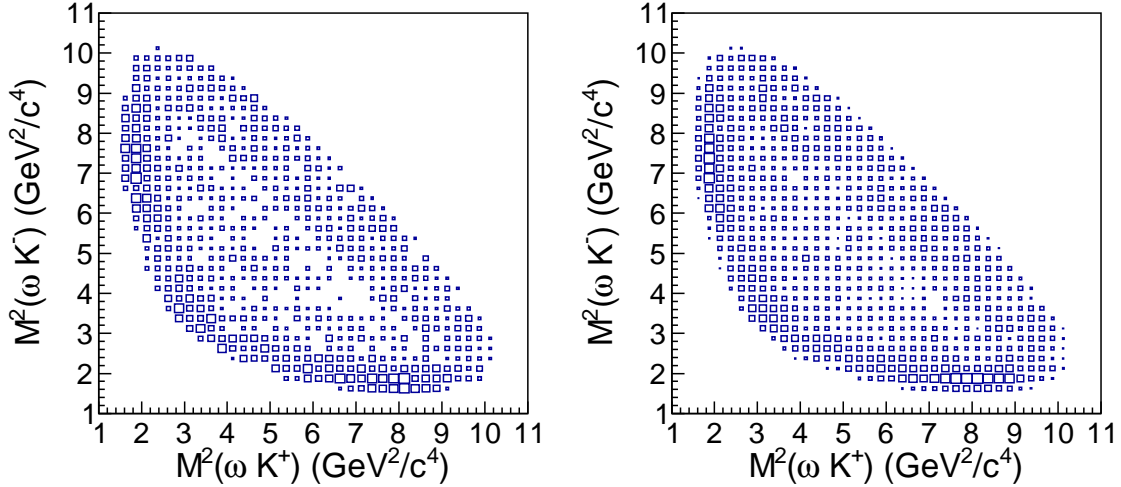}
\caption{The Dalitz plots of the data (left) and exclusive MC simulation (right) with
  the BODY3 generator for events in the region of $0.772< m_{\pi^+\pi^-\pi^0}<0.792~\gevcc$.}
\label{fig:dalitz}
\end{figure}
\begin{figure}
\includegraphics[width=5.0in]{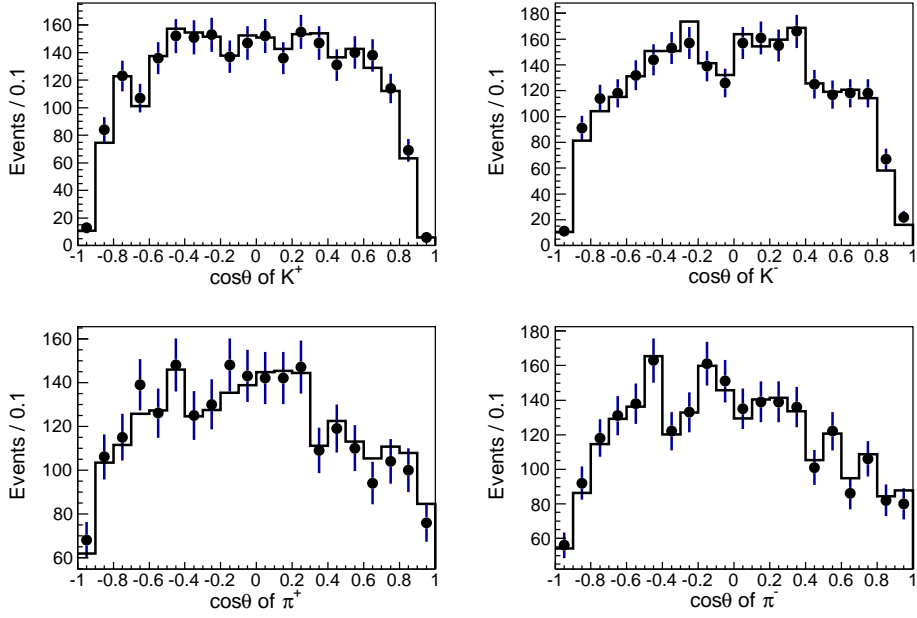}
\caption{Comparison of various $\cos\theta$ distributions for
  $\psp\to\omega K^+K^-$ candidates of data and signal MC with BODY3
  generator for events in the region of $0.772< m_{\pi^+\pi^-\pi^0}<0.792~\gevcc$.
  The $\cos\theta$ of $K^+$ and $K^-$ is measured in the center-of-mass frame of $\psp$,
  and that of $\pi^+$ and $\pi^-$ is of $\omega$.
  Dots with error
  bars are data, and the histograms are the signal MC with the BODY3
  generators.}
\label{fig:angle}
\end{figure}
\begin{figure}
\includegraphics[width=5.0in]{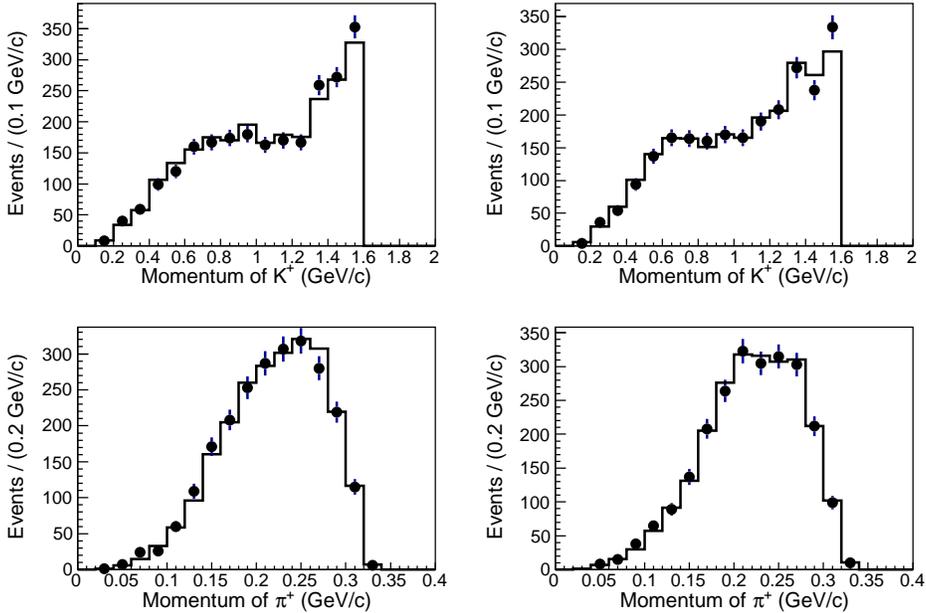}
\caption{Comparison of various momentum distributions for
  $\psp\to\omega K^+K^-$ candidates of data and signal MC with both
  BODY3 generator for events in the region of $0.772< m_{\pi^+\pi^-\pi^0}<0.792~\gevcc$. The dots
  with error bars are data, and the histogram is the signal MC with the
  BODY3 generators.}
\label{fig:mom}
\end{figure}

In this analysis there are two three-body
decay chains, i.e. $\psp \to \omega K^+ K^-$ and $\omega \to \pi^+ \pi^- \pi^0$, so two
BODY3 generators are applied in sequence to simulate the whole process.  The events in the
region of $0.772< m_{\pi^+\pi^-\pi^0}<0.792~\gevcc$ are used to give the probability
distribution function as most of $\omega$ candidates are in this region.  For the energy
points at $3.686~\gev$ and $3.65~\gev$, two different sets of data are used as input to
the event generation.  Using $3.686~\gev$ as an example, the comparison between data and
MC of Dalitz plots, momentum and angular distributions are displayed in
Figs.~\ref{fig:dalitz},~\ref{fig:angle} and~\ref{fig:mom}, respectively. The MC simulation
matches the data well in every distribution, so the determination of the signal efficiency
should be reliable.
While backgrounds have not been subtracted in the BODY3 simulation, their effect will be considered as a systematic uncertainty (see Section IV).

\begin{figure}
\includegraphics[width=5.0in]{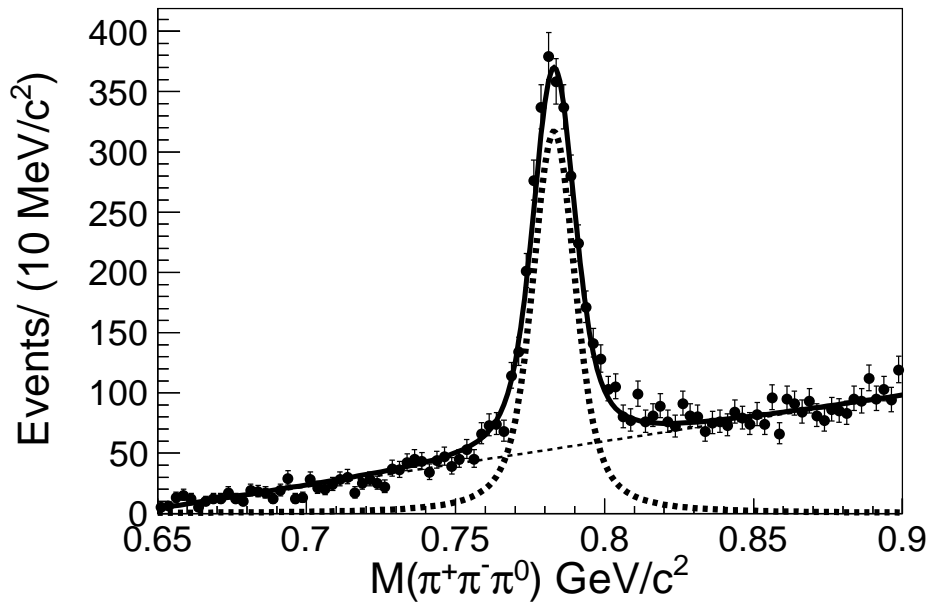}
\caption{Fit to data to obtain the yields at $3.686$ GeV. The
  solid line is the total fit result, the dots with error bars are
  data, the bold dashed line is the signal shape, and the thin dashed
  line is the background.}
\label{fig:08}
\end{figure}

Neglecting possible interference between resonance and non-resonance
processes, the $\omega K^+ K^-$ yield is
obtained by fitting the $m_{\pi^+ \pi^- \pi^0}$ distribution. The
signal shape is described by a smeared Breit-Wigner function,
i.e. $\mathrm{BW} \otimes \mathrm{Gauss}$, where the $\sigma$ of the
Gaussian describes the resolution and the width of the $\omega$
is fixed at $8.49~\mathrm{MeV}/\mathrm{c}^2$ according to the
world average value~\cite{Beringer:1900zz}. The background is described by a linear function.
For the $\psp$ data sample, the number
of observed signal events is $2781\pm68$, and the fit is shown in
Fig.~\ref{fig:08}.  From the fit, the mass of $\omega$ is
$783.1\pm0.2~\mathrm{MeV}/\mathrm{c}^2$, the resolution is $(5.05\pm0.28)~\mathrm{MeV}/\mathrm{c}^2$
and the goodness of the fit is $\chi^2/ndf=107/95=1.13$.

\begin{figure}
\includegraphics[width=5.0in]{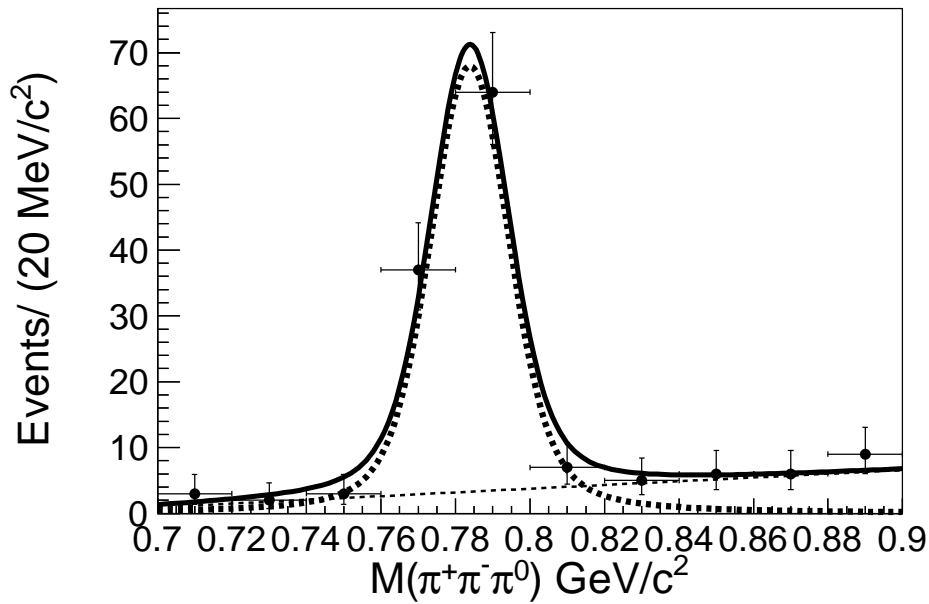}
\caption{Fit to $3.65$ GeV data to obtain the continuum
  yield. The solid line is the total fit result, the dots with error
  bars are data, the bold dashed line is the signal shape, and the
  thin dashed line is the background.}
\label{fig:cont}
\end{figure}

A similar event selection and fit method are applied to the $44.49~\pbv$
$e^+e^-$ data sample collected at $3.65$ GeV, in
which $100 \pm 11$ signal events are observed, and the fit is shown in
Fig.~\ref{fig:cont}. The fit result gives the
$\omega$ mass $782.0\pm 0.9~\mathrm{MeV}/\mathrm{c}^2$, the resolution is $4.4\pm1.6~\mathrm{MeV}/\mathrm{c}^2$
and the goodness of fit $\chi^2/ndf=4.20/4=1.05$.

Under the assumption that interference between $\psp$ decay and
continuum production of the same final state is absent, the branching
fraction of $\psi(3686) \to \omega K^+K^-$ is determined by the
formula
\begin{equation}
\mathcal{B}(\psp \to\omega K^+K^-) =
 \frac{N_{3.686}/\epsilon_{3.686}- f_{c} \cdot N_{3.65}/\epsilon_{3.65}}
{\mathcal{B}(\omega\to\pi^0\pi^+\pi^-)\times \mathcal{B}(\pi^0\to\gamma\gamma)\times
 N_{\psp}} \; ,
\label{eq:br}
\end{equation}
where $\mathcal{B}(\omega\to\pi^0\pi^+\pi^-) = 0.892\pm 0.007$ is the
branching fraction of $\omega \to \pi^+ \pi^-
\pi^0$~\cite{Beringer:1900zz}, $\mathcal{B}(\pi^0\to\gamma\gamma) =
0.98823 \pm 0.00034$ is the branching fraction of $\pi^0 \to \gamma
\gamma$~\cite{Beringer:1900zz}, $N_{\psp} = (106.41\pm0.86)\times10^6$
is the number of $\psp$ events~\cite{Ablikim:2012pj}, and the scaling
factor $f_c = 3.677$ is determined from the luminosities and continuum
hadronic cross sections of the two data samples used in this
paper~\cite{Ablikim:2012pj}. The efficiencies $\epsilon_{3.686} =
16.9\%$ at $3.686$ GeV and $\epsilon_{3.65} = 20.7\%$ at $3.65$
GeV. $N_{3.686}$ and $N_{3.65}$ are the numbers of events observed in
the $3.686$ GeV and $3.65$ GeV data samples, respectively. Thus
$\mathcal{B}(\psp\to\omega K^+K^-)$ is determined to be $(1.56 \pm
0.04)\times 10^{-4}$, where the uncertainty is only statistical.

\section{Systematic uncertainties}

The tracking efficiencies of $K$ and $\pi$ have been studied
with data samples for the clean processes $J/\psi\to K^0_S K^\pm\pi^\mp + c.c.$, $K^0_S\to
\pi^+\pi^-$ and $\psi(3686)\to \pi^+ \pi^- J/\psi$, $J/\psi\to l^+l^-$,
respectively. The difference of charged track efficiencies between
data and MC simulated events is $1\%$ per track~\cite{Ablikim:2013gd, Ablikim:2012pg}. Therefore
$4\%$ is taken as the total uncertainty in tracking efficiency due to
four charged particles in the final states.

The photon reconstruction efficiency has been studied via the
processes $J/\psi\to \rho^0 \pi^0$, $\rho^0 \to \pi^+ \pi^-$, $\pi^0
\to \gamma \gamma$, and $1\%$ is taken as the systematic uncertainty
of photon reconstruction~\cite{Ablikim:2010zn}. The total systematic
uncertainty of photon reconstruction here is $2\%$ since the final
states has two photons.

The PID systematic uncertainty is $1\%$ for each charged particle,
determined from $J/\psi \to \pi^+\pi^-\pi^0$ and $K^+K^-\pi^0$~\cite{Ablikim:2011kv}, so the total PID systematic uncertainty
is $4\%$ due to four charged tracks in the final states.

The systematic uncertainty of the kinematic fit is estimated by using the
method described in Ref.~\cite{Ablikim:2012pg}, where
helix parameter corrections
corresponding to the difference between data
and MC is made, and the difference of the efficiencies with and
without this correction is taken as the systematic uncertainty. In
this analysis, the efficiency changes from $17.5\%$ to $17.3\%$ after
this correction, so we take $1.1\%$ as the systematic uncertainty of
the kinematic fit.

The systematic uncertainty of the background shape is estimated by
checking the results with different background shapes, and the maximum
difference is quoted as its uncertainty.  We used 1st, 2nd and 3rd
order polynomial backgrounds, and varied the fit region of
$m_{\pi^+\pi^-\pi^0}$ from $[0.65$ GeV/$c^2$, $0.90$ GeV/$c^2$$]$ to
$[0.60\ \mathrm{GeV}/c^2,\ 0.95\ \mathrm{GeV}/c^2]$.  The biggest
signal, $2896 \pm 72$, in the $3.686\ \mathrm{GeV}/c^2$ data sample is
obtained using a second order polynomial background shape and a fit
region $[0.64\ \mathrm{GeV}/c^2,\ 0.91\ \mathrm{GeV}/c^2]$; the
difference with the nominal result yields a $3\%$ systematic
uncertainty.

The systematic uncertainty with the BODY3 generator is composed of
three parts. The first one is attributed to the limited statistics of
the data sample, which is used as input to construct
the data simulated by the BODY3 generator.
The second one is attributed to the binning
method. The third one is from the remaining backgrounds. The first
uncertainty is obtained directly. The second one is obtained by
varying the binning.  The third one is estimated using inclusive MC as
input and determining the change of efficiency with and without
background . Combining the uncertainties of these three parts, the
final uncertainty from the BODY3 generator is $1.3\%$.

The trigger efficiency is very high due to four charged tracks and two
photons in the final states~\cite{Berger:2010my}, and the systematic
uncertainty of the trigger efficiency can be neglected in this
analysis.

The number of $\psp$ events is $(106.41\pm0.86)\times10^6$, which is
determined using $\psp\to \mathrm{hadrons}$~\cite{Ablikim:2012pj}. The
uncertainty of $f_c$ is small, 0.2\%~\cite{Ablikim:2012pj}, and yields
a negligible systematic error on the branching ratio.

The systematic uncertainty of the $\pi^0$ selection is estimated by
removing
the requirement of $0.11~\gevcc < M_{\gamma\gamma}<0.15~\gevcc$ in event selection.
The difference of the result, $0.6\%$ is taken as the
systematic uncertainty.

Table~\ref{table:totalerror} compiles all sources of systematic uncertainties
in the measurement of the branching fractions,
and the total systematic uncertainty is
$7.0\%$, which is obtained by adding the uncertainties in quadrature.

\begin{table}[!hbp]
\caption{Summary of systematic uncertainties.}
\label{table:totalerror}
\begin{tabular}{|l|c|}
\hline
\hline
\bf{Source of uncertainty} & \bf{Uncertainty} \\
\hline
MDC tracking & 4.0\% \\
\hline
PID & 4.0\% \\
\hline
Photons & 2.0\% \\
\hline
Kinematic fit & 1.1\% \\
\hline
Background shape & 3.0\% \\
\hline
$N_{\psp}$ & 0.8\% \\
\hline
$f_c$ & - \\
\hline
BODY3 & 1.3\% \\
\hline
Trigger & -  \\
\hline
Resolution of $\pi^0$ & 0.6 \\
\hline
Total   &  $7.0\%$ \\
\hline \hline
\end{tabular}
\end{table}

\section{Summary}
In this paper, the branching fraction of $\psi(3686)\to\omega K^+K^-$ is
measured to be $(1.56\pm 0.04 \pm 0.11) \times10^{-4}$. The comparison
with previous results is displayed in Table~\ref{table:compare}, and
our result is the most precise measurement to date. There is a 1.7 $\sigma$ (statistical and systematic)
difference between the BESIII and BESII measurements of
$\mathcal{B}(\psi(3686)\to\omega K^+K^-)$, and the precision of the
BESIII measurement is greatly improved compared to BESII. Part of the improvement is
attributed to the determination of the continuum contribution. With
the much larger integrated luminosity ($44.49\ \mathrm{pb}^{-1}$ and $6.4\
\mathrm{pb}^{-1}$ for BESIII and BESII) and much higher reconstruction
efficiency ($20.7\%$ and $2.4\%$ for BESIII and BESII) respectively, the
contribution from the continuum process has been determined with much
higher precision.


\begin{table}[!hbp]
\caption{Comparison of our result with previous measurements of the branching fraction
  $\psp\to\omega K^+K^-$ and to the world average from particle data group (PDG).}
\label{table:compare}
\begin{tabular}{|l|l|}
\hline
\hline
Branching fraction    &     Source     \\
\hline
$(1.56\pm 0.04 \pm 0.11) \times10^{-4}$  & this analysis \\
\hline
$(2.38\pm0.37\pm0.29)\times10^{-4}$ & BESII~\cite{Ablikim:2005ju} \\
\hline
$(1.9\pm0.3\pm0.3)\times10^{-4}$ & CLEO~\cite{Briere:2005rc} \\
\hline
$(1.5\pm0.3\pm0.2)\times10^{-4}$ & BES~\cite{Bai:2002yn} \\
\hline
$(1.85\pm0.25)\times10^{-4}$ & PDG~\cite{Beringer:1900zz} \\
\hline
\hline
\end{tabular}
\end{table}


From the world average value~\cite{Beringer:1900zz}, the branching fraction of
$J/\psi\to\omega K\bar K$ is $(1.70\pm0.32)\times10^{-3}$, and assuming on basis of
isospin symmetry that one half is charged kaons, the branching fraction of
$J/\psi\to\omega K^+K^-$ is $(0.85\pm0.16)\times10^{-3}$. Therefore,
$Q\approx(18.4\pm3.7)\%$, which is smaller than the previous result $(21.8 \pm 5.0)\%$
based on the branching ratio of $\psi(3686)\to\omega K\bar K$ of the world average
value. With the improvement on the measurement of $\psp \to \omega K^+ K^-$, the
uncertainty on $Q$ now mainly stems from $J/\psi \to \omega K^+ K^-$, and a measurement of
the branching fraction of $J/\psi\to\omega K^+K^-$ with at least same precision is needed
in order to establish a significant deviation from 12\% rule.

\section{Acknowledgments}
The BESIII collaboration thanks the staff of BEPCII and the computing center for their
strong support. This work is supported in part by Joint Funds of the National Natural
Science Foundation of China under Contract Nos. U1232109, 11179020; the Ministry
of Science and Technology of China under Contract No. 2009CB825200; National Natural
Science Foundation of China (NSFC) under Contracts Nos. 10625524, 10821063, 10825524,
10835001, 10935007, 11125525, 11235011, 11079008, 11179007, 11105101, 11205117, 11375221; the Chinese Academy of
Sciences (CAS) Large-Scale Scientific Facility Program; CAS under Contracts
Nos. KJCX2-YW-N29, KJCX2-YW-N45; 100 Talents Program of CAS; German Research Foundation
DFG under Contract No. Collaborative Research Center CRC-1044; Istituto Nazionale di
Fisica Nucleare, Italy; Ministry of Development of Turkey under Contract
No. DPT2006K-120470; U. S. Department of Energy under Contracts Nos. DE-FG02-04ER41291,
DE-FG02-05ER41374, DE-FG02-94ER40823, DESC0010118; U.S. National Science Foundation;
University of Groningen (RuG) and the Helmholtzzentrum fuer Schwerionenforschung GmbH
(GSI), Darmstadt; WCU Program of National Research Foundation of Korea under Contract
No. R32-2008-000-10155-0.

\end{document}